\documentclass[12pt]{article}

\usepackage{cite}
\usepackage{graphics}
\usepackage{graphicx}
\usepackage{subfig}
\usepackage{xcolor}
\usepackage{url} 

\title{Investigation of the relationship between code change set n-grams and change in energy consumption}
\date{
\{romansky, abram.hindle\}@ualberta.ca}
\author{Stephen Romansky\\{\small Supervisor: Abram Hindle}}

\begin{document}
\maketitle

\begin{abstract}
The amount of software running on mobile devices is constantly growing as
consumers and industry purchase more battery powered devices. On the other
hand, tools that provide developers with feedback on how their software
changes affect battery life are not widely available.
This work employs Green Mining, the study of the relationship between energy
consumption and software changesets, and n-gram language models to evaluate if
source code changeset perplexity correlates with change in energy consumption.
A correlation between perplexity and change in energy consumption would permit the
development of a tool that predicts the impact a code changeset may have on a
software applications energy consumption.
The case study results show that there is weak to no correlation between
cross entropy and change in energy consumption. Therefore, future areas
of investigation are proposed.

\end{abstract}

\section{Introduction}
Consumer's and industry's increasing usage of battery limited mobile
devices --- whether they are
tablets, laptop computers, or mobile phones ---
provide significant motivation for improving energy efficiency due to
the benefit of longer battery life.
Longer mobile battery life, without impairing the users experience, is a
selling point for industry producers.
Low battery life limits the availability of a device to industrial workers.
In some cases low-battery notifications, or lack of power to a device, can
interrupt industrial workers in their work flow. Thus, resulting in a loss of
work place efficiency.
This work looks to improve mobile battery life by investigating software
change and its impact on energy consumption.

Many software applications run on feature rich mobile devices.
The implementation of how a software application interacts with a given
hardware feature may have a negative impact on energy consumption of the
device running the software.
Therefore, software development recommendations~\cite{iOSPG, win7perf} have
been created by industry to aid developers with mobile application development.
Unfortunately, recommendations do not give developers feedback on whether or
not hardware usage in their mobile applications is efficient.
Furthermore, recommendations do not give developers awareness of how their
code changesets will influence their mobile application's energy consumption.
This work explores the question ``What features of software changesets can be
used to give mobile application developers feedback on how their software
changesets will impact energy consumption?''

Researchers have searched for what, whether software, libraries,
hardware, or behaviors, consumes significant amounts of energy in
mobile devices.
Steigerwald~\emph{et al.}~\cite{steigerwald2011developing} points out that software-hardware interactions can cause energy
consumption inefficiencies depending on the software's computation, data, and
idling efficiencies.
Hardware profiling~\cite{carroll2010analysis} of mobile devices was also
explored to try and improve power management for an increase in energy
efficiency.
These researchers have not investigated how code changesets can impact a
software applications energy consumption.
Therefore,
this work raises the research question ``Does the cross-entropy of a change to
a software application's codebase correlate with a change in the software's
energy consumption?''

\section{Prior Work}

Firstly, we must introduce what power and energy are.
A watt is a unit from the international system of units (SI). Watts express
energy consumption over time or joules per time. Joules are another SI unit;
joules describe a unit of energy.

Green Mining~\cite{hindle2012greenA, hindle2012greenB} was proposed to study
the relationship of software change and the influence it has on energy
consumption.
Green Mining shows that software commits can introduce change in a
software's energy consumption~\cite{hindle2012greenA}.
To collect data, on how energy consumption changes with respect to new software
revisions, a Green Mining practitioner constructs a set of tests for a given
software product.
The tests are run repeatedly on a set of revisions for the software product
under test while energy consumption metrics are recorded.
The result of this process is a historic energy profile for the given
application.
The energy profile can be modeled to show how the application evolved over
time with respect to energy consumption.
Motivation to further this area of study comes from the potential to create
development tools that aid developers to create more energy efficient software.
Development aids are not readily available for developers to use at the
present time.

The process of Green Mining is tedious and consumes a significant amounts of
human time. To
make the process of Green Mining easier for practitioners the Green
Miner~\cite{hindle2014green} was innovated. The Green Miner provides a testbed
for running software energy consumption tests reliably and repeatedly. This
saves time for researchers and reduces the error rate humans could introduce
during data collection.

The Green Miner implementation~\cite{hindle2014green} uses kits that can
measure the energy consumption of Android mobile phones. This is done using
an Arduino, a Raspberry Pi, an INA 219 IC, and an Android phone to create a
kit. A power supply passes through the INA 219 to the battery contacts of
the mobile device. Energy consumption data from the INA 219 is relayed to the
Raspberry Pi until testing has been completed.
A web interface allows tests to be scheduled and run on the mobile
phones. Once the tests have completed, the Raspberry Pi will create a
tarball that the web interface will collect and store for a researcher.


This work converts Green Mined software application codebases and changesets
into n-gram language models (LM). This is because codebases provide corpora
which tokenized changesets can be compared against to calculate the
cross-entropy of changesets.
An n-gram is $n$ continuous subsequent items of a set. For
instance, in the previous sentence a word 3-gram would be ``An n-gram is'' or
``is a subset''.
N-grams are used to construct probability models by calculating the probability
of the $n$th item occurring given the $n-1$ prior items.
For example, the probability we see the word ``is'' in the prior
example given that ``An n-gram'' have been seen could be calculated.

To create the LMs, MIT LM (MITLM~\cite{hsu2008iterative}) toolkit is used.
The toolkit lets researchers easily compare two corpuses using perplexity.
Perplexity is how well two corpuses fit each other. If a high perplexity is
calculated
then the two language models derived from the corpuses do not fit each other
well.

\section{Related Work}
Green Mining practitioners have investigated software changesets
for features that correlate with change in software energy consumption.
One such work~\cite{hindle2012greenB} shows that lines of code in a
changeset do not correlate with change in energy consumption.
In another study~\cite{hindle2012greenA}, the object oriented metrics Number of
Children (NOC) and Depth of Inheritance Tree (DIT) were found to have a
rank-correlation with mean energy consumption.
Aggarwal~\emph{et al.}~\cite{aggarwal2014cascon} show a relationship can exist
between change in syscall profiles and energy profiles.
There are still many features of software changesets to be investigated.

Muthukumarasamy~\cite{muthukumarasamy2010extraction} constructed n-gram models
from assembly code. The n-gram model has weights for the execution time and
energy consumption associated with itself.
The work~\cite{muthukumarasamy2010extraction} demonstrates how to use the
weighted n-gram model to predict the execution time and energy consumption of
a software application by using static prediction analysis.
This method only looks at a single version of the software application instead
of investigating the change in execution time and energy consumption that can
be introduced by new revisions of the software.

Researchers have investigated the applicability of n-gram models to software
engineering problems in prior works.
Abou-Assleh~\emph{et al.}~\cite{abou2004n} demonstrates the usage of n-gram
models in detecting malicious code.
Campbell~\emph{et al.}~\cite{campbell2014syntax} utilizes n-gram models to
improve source code error detection at a software's compile time.
The investigation done in this work checks whether or not n-gram language
models are useful in predicting change in software energy consumption.
None of these applications have been to software changesets.

\section{Method}
To partially answer the research question ``which software changeset feature
correlates with change in energy consumption?'', the perplexity of language
models, which are derived from source code, are investigated. If perplexity
of a software changeset, with
respect to a codebase, correlates with change in energy consumption then this
software feature could be used to predict the impact a changeset may have on
energy
consumption. This method checks if perplexity and change in energy consumption
correlate using three different approaches.

An energy profile is constructed from an Android application.
Figure \ref{fig:fennec} shows an energy profile for the Fennec application.
The energy
profile
provides a view of historic energy consumption for the selected application.
In addition, the energy profile provides information to calculate the change in
energy consumption each revision introduces. The energy change can be
calculated by
taking two consecutive revisions from the energy profile and subtracting the
foremost revision's energy consumption from the latter revision's energy
consumption.

The energy profile provides a list of software revisions that have been
studied. Given that this work is interested in finding the perplexity of
code changesets, and that the energy profile does not have code changesets in
it, the changesets need to be found.
Using the revision list, and the software repository the revisions are
associated with, it is possible to generate all of the changesets corresponding
to the change in energy consumption values from the energy profile.
Thus, there is a set of code
changesets and a list of the associated energy consumption values that have
been collected. The collection of data permits perplexity to be investigated.

\paragraph{Part 1:}
Code changesets and codebase of an Android application are converted to
language
models by first removing non-\texttt{C++} code, and tokenizing the remaining
\texttt{C++} code. A set of tokenized changesets is created for lines added
and lines removed, of \texttt{C++} code.
The MIT Language Model (\texttt{MITLM}) toolkit is applied to a tokenized
codebase and tokenized changesets to calculate the perplexity of code
changesets with respect to the codebase.

\paragraph{Part 2:}
This work also investigates generating language models from multiple
codebases, the immediate past, instead of using a single static codebase as in part 1. Each
revision of the
software application is checked out from the application's version control
system and tokenized with respect to \texttt{C++}.
For each tokenized codebase the following subsequent 35 software changesets
have their perplexities calculated. The perplexities are then correlated
with the 35 energy consumption values associated with the changesets, for each
of the codebases.

\paragraph{Part 3:}
The final approach 
groups changesets based on change in
energy consumption to
 to check whether or not perplexity and change in
energy consumption correlate, 
The absolute value is calculated for each change in energy consumption
value in the software applications energy profile.
The data is partitioned into three sets, the low change set, the medium change
set, and the high change set.
The low change set contains software revisions that have a change in absolute
energy consumption from and including -1 standard deviations to 0 standard
deviations.
The medium change set contains revisions with a change in absolute energy
consumption from and including 0 standard deviations to 1 standard deviation.
The high change set contains revisions with a change in absolute energy greater
than 1 standard deviation.
The partitions can then be plotted using box plots such that their median values, and
ranges, can be compared. The comparison is to check if the groupings have any
correlation with perplexity similar to part 1 and 2.

\section{Case Study}
The Android Fennec web browser application was mined in prior
work~\cite{hindle2014green}. The process for extracting a historic energy profile,
as outlined in Green Mining~\cite{hindle2012greenA, hindle2012greenB}, involves identifying a software
candidate that has many software revisions available for test, writing energy
consumption tests for the chosen software application, and repeatedly running
the tests while collecting energy consumption data on the software application.

The process of extracting an energy profile manually is tedious. Therefore, the
Green Miner~\cite{hindle2014green}, an automated test-bed, is used to ease data
collection. The Green Miner implementation consists of instrumented Android
mobile devices, and a web interface. The web interface allows a researcher to
schedule written tests on the mobile devices. One testing has completed the
test-bed will aggregate collected energy consumption data from the
instrumented devices automatically for the researcher.

\begin{figure}
  \centering
  \includegraphics[scale=0.50]{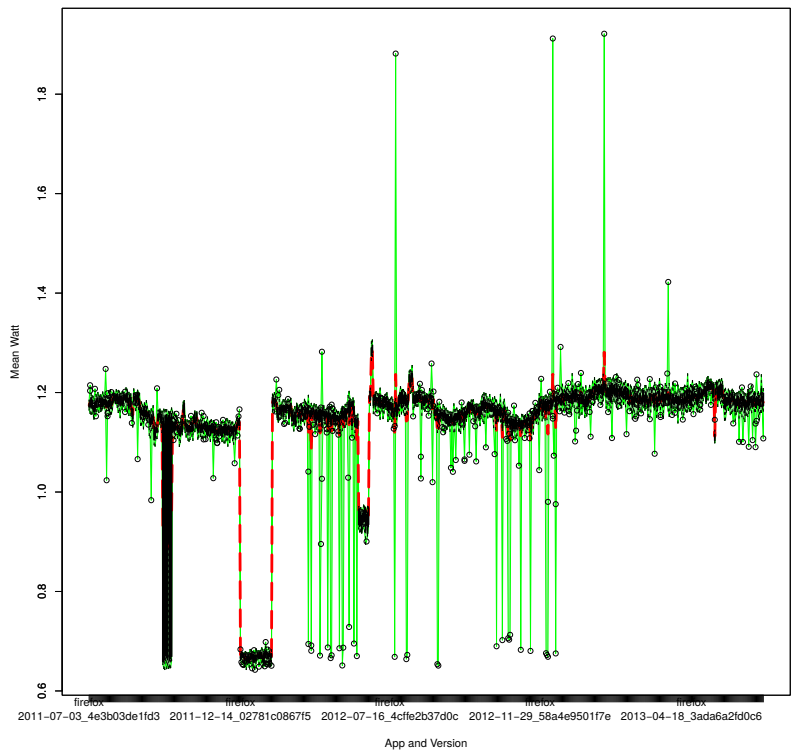}
  \caption{Fennec Web Browser Energy Consumption for each Software Revision}
  \label{fig:fennec}
\end{figure}

The Android Fennec web browser had its energy consumption tested by simulating a
user reading a web page. The test was repeated on 686 revisions, more than 10
times each, of the Fennec application.
Figure \ref{fig:fennec} shows the energy consumption measured in mean watts for
each tested revision of the Fennec application.

The Fennec application was versioned using the mercurial version control
system at the time of writing. To generate a changeset using mercurial,
on linux, the following command can be used given two revisions:
\begin{verbatim}
hg diff -r <revision1>:<revision2>
\end{verbatim}
The output of the command is the changesets between the two specified
revisions. A script was used to iterate through all consecutive revisions, in
the energy profile, and call the mercurial command to generate changesets.
An additional script was used to create sets of added lines of C++ code and sets of removed
lines of C++ code from the source changesets. The added and removed lines of
C++ were tokenized\footnote{Arash Partow of partow.net wrote a C++ lexer
toolkit that was modified by the author, Stephen Romansky} and saved for later processing.

\begin{figure}
  \centering
  \includegraphics[scale=0.50]{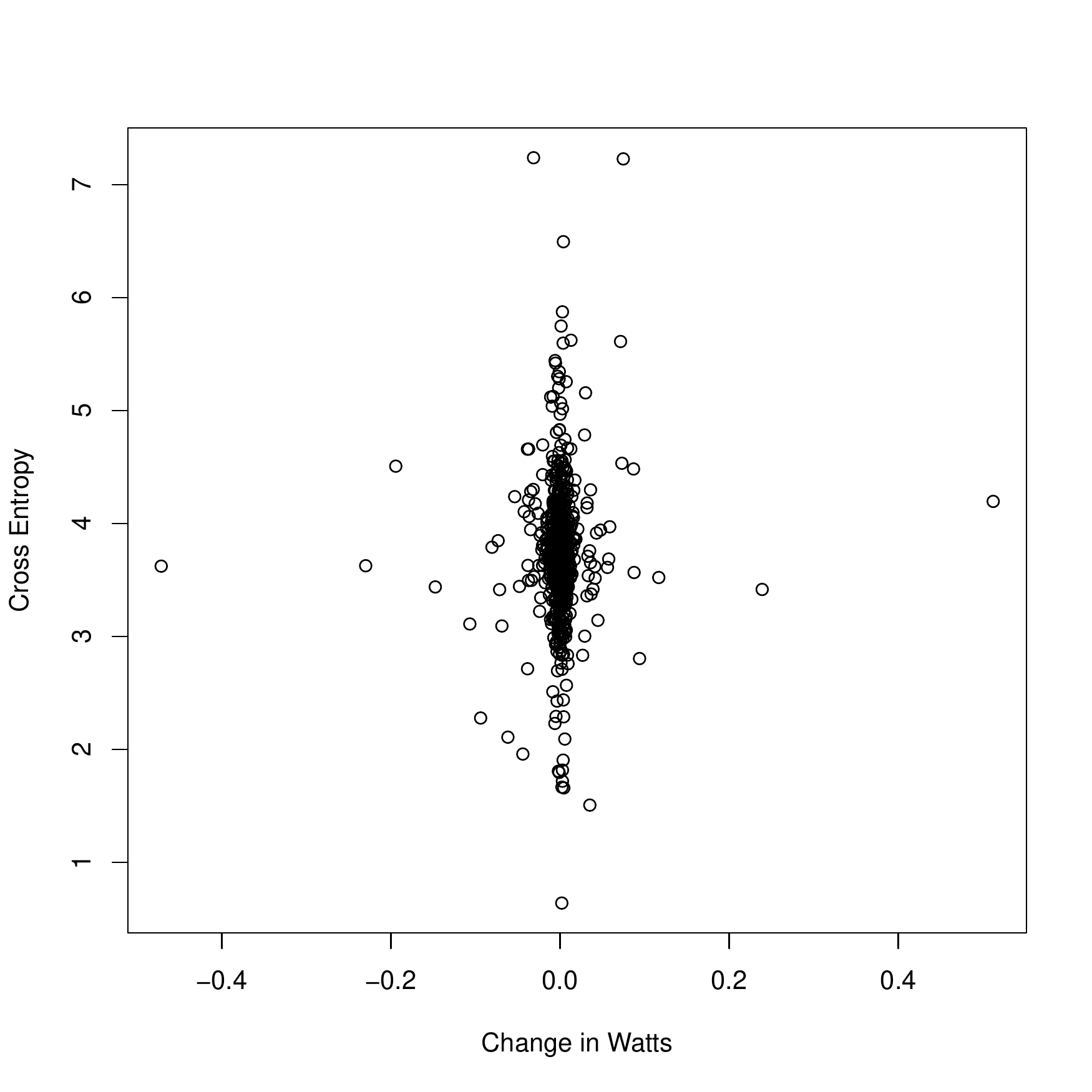}
  \caption{Cross Entropy versus Change in Watts per software changesets}
  \label{fig:addedLineToken}  
\end{figure}

\begin{figure}
  \centering
  \includegraphics[scale=0.50]{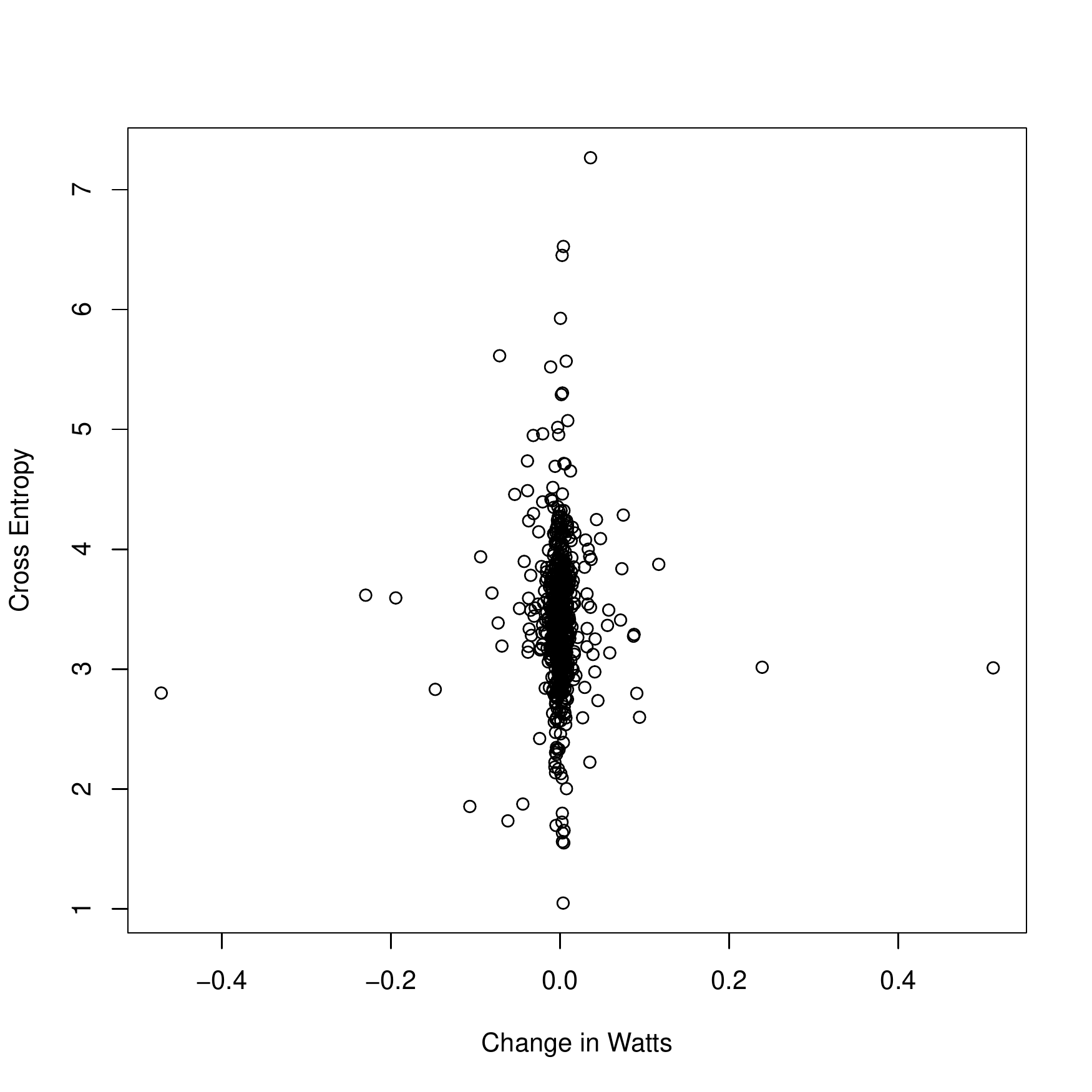}
  \caption{Cross Entropy versus Change in Watts per software changesets}
  \label{fig:removedLineToken} 
\end{figure}

\paragraph{Part 1:}
To create a codebase to calculate perplexities with version 4e3b03de1fd3 of
the Fennec application was checked out from the software repository. All of
the C++ code from the repository was tokenized from the codebase.
The \texttt{MITLM} utility was then used, with a 3-grams and modified
Kneser-Ney smoothing, to calculate the perplexity of the added C++ line
changesets and the removed C++ line changesets with respect to the tokenized
codebase. Figure \ref{fig:addedLineToken} shows cross entropies of tokenized
added line changesets, with respect to the 4e3b03de1fd3 codebase, versus
change in energy consumption. Figure \ref{fig:removedLineToken} shows cross entropies of
tokenized removed line changesets, with respect to the 4e3b03de1fd3 codebase,
versus change in energy consumption. The cross entropy outliers have been
removed from Figure \ref{fig:addedLineToken} and \ref{fig:removedLineToken}.

\begin{figure}
  \centering  
  \includegraphics[scale=0.50]{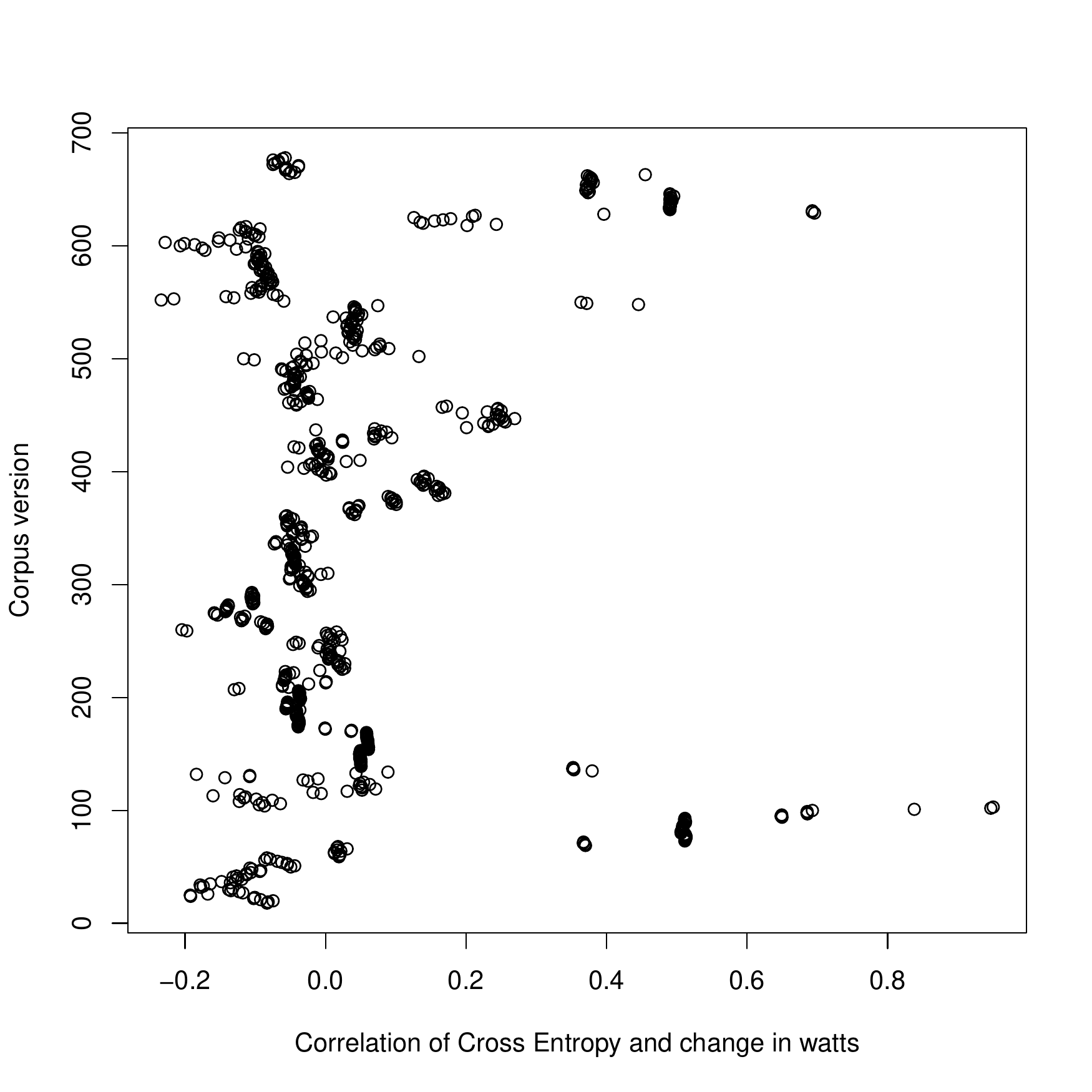}
  \caption{Correlation of Cross Entropy and change in watts for each codebase corpus using added line changesets}
  \label{fig:addedMovingCorp}  
\end{figure}

\begin{figure}
  \centering
  \includegraphics[scale=0.50]{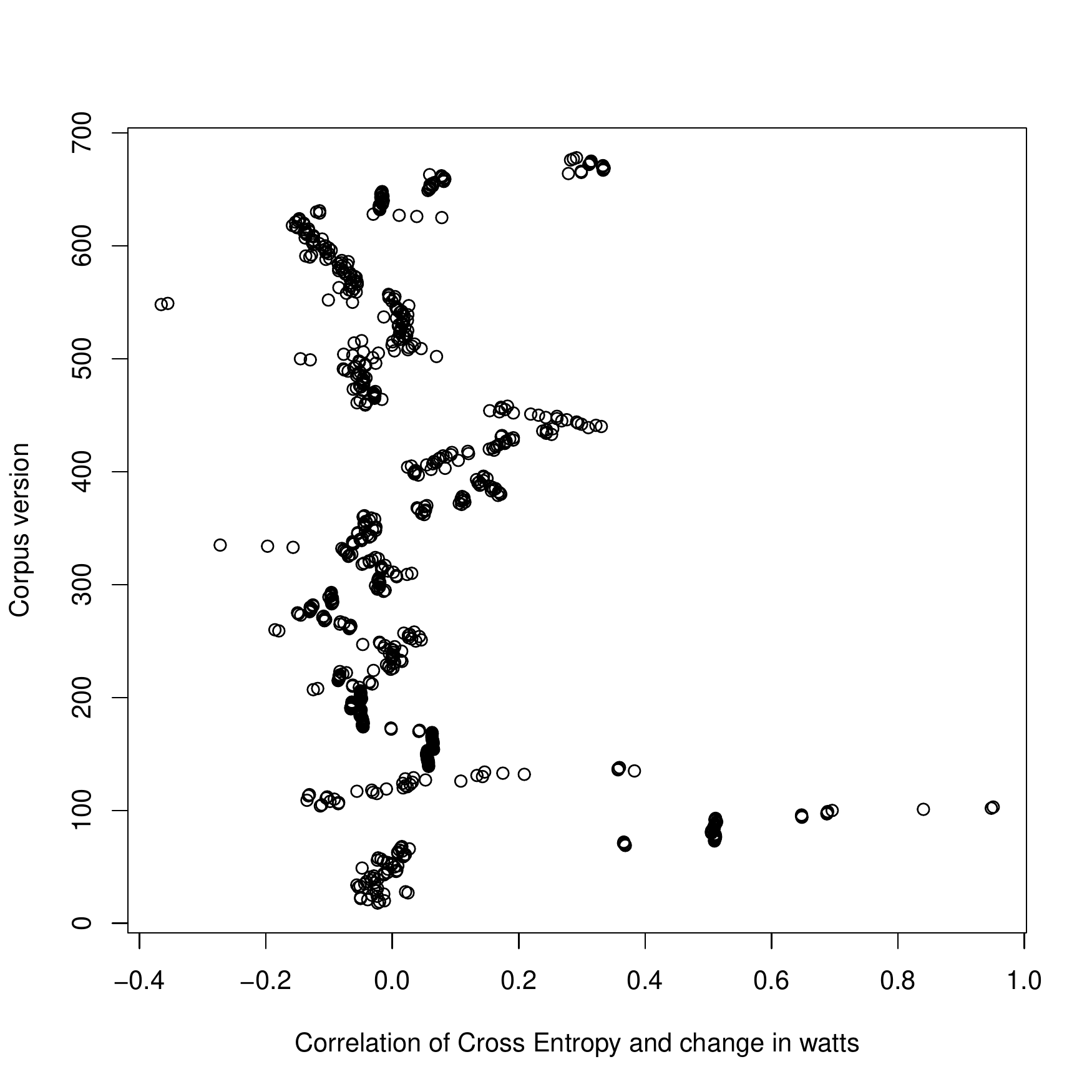}
  \caption{Correlation of Cross Entropy and change in watts for each codebase corpus using removed line changesets}
  \label{fig:removedMovingCorp}  
\end{figure}

\paragraph{Part 2:}
Each revision in the energy profile was used to create a lexed codebase
corpus. A
script was used with the mercurial tool to generate a lexed corpus for
each version of the Fennec repository in the energy profile. Each corpus
was then used to calculate the cross entropy of the following 35
revisions. The points in Figure \ref{fig:addedMovingCorp} and
\ref{fig:removedMovingCorp} are the correlation values for each lexed
corpora's 35 code changeset cross entropy and change in energy consumption
values.

\begin{figure}
  \centering
  \includegraphics[scale=0.50]{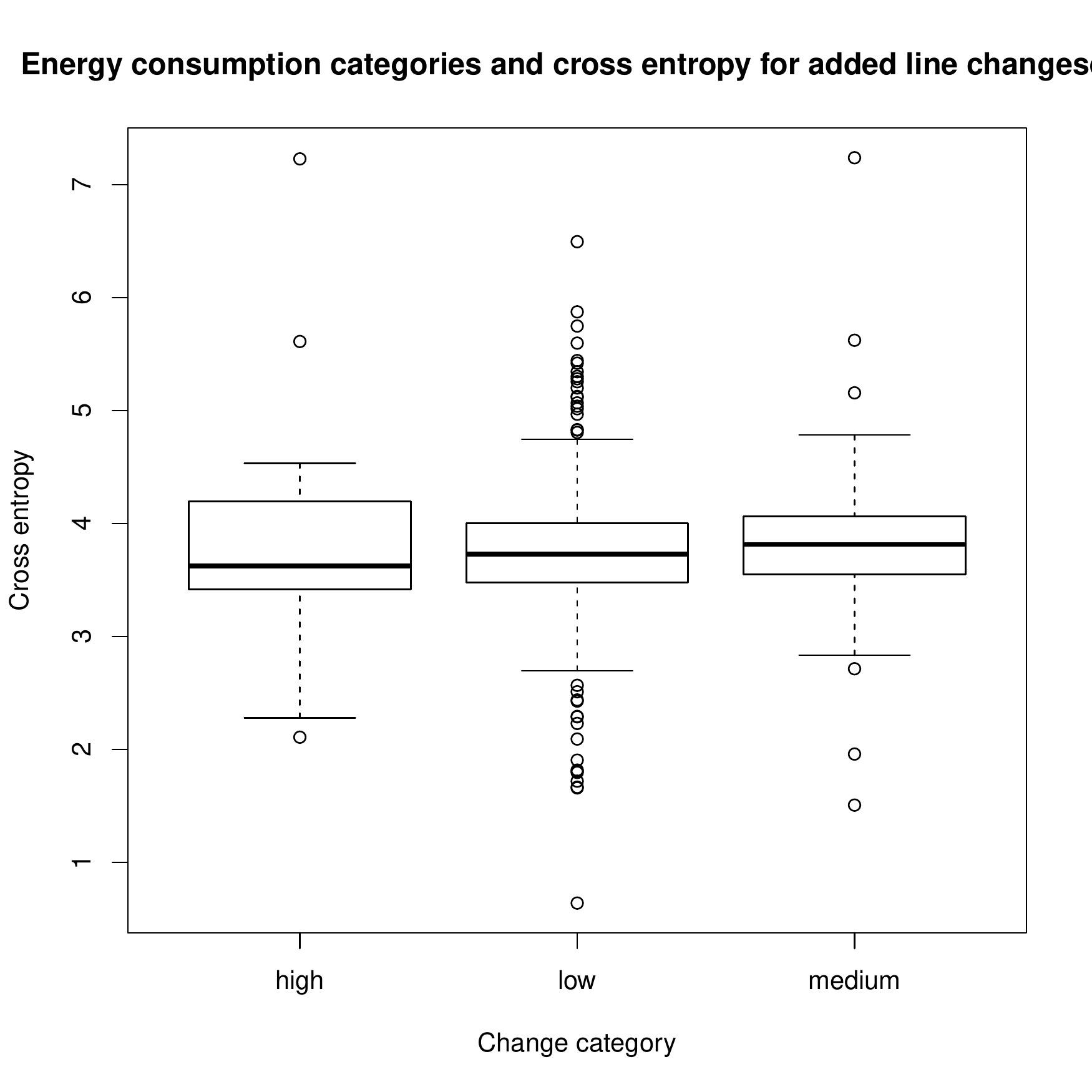}
  \caption{Changeset Cross Entropy Grouped by Change in Energy Consumption of added line changesets}
  \label{fig:threeGroupAdded}
\end{figure}

\begin{figure}
  \centering
  \includegraphics[scale=0.50]{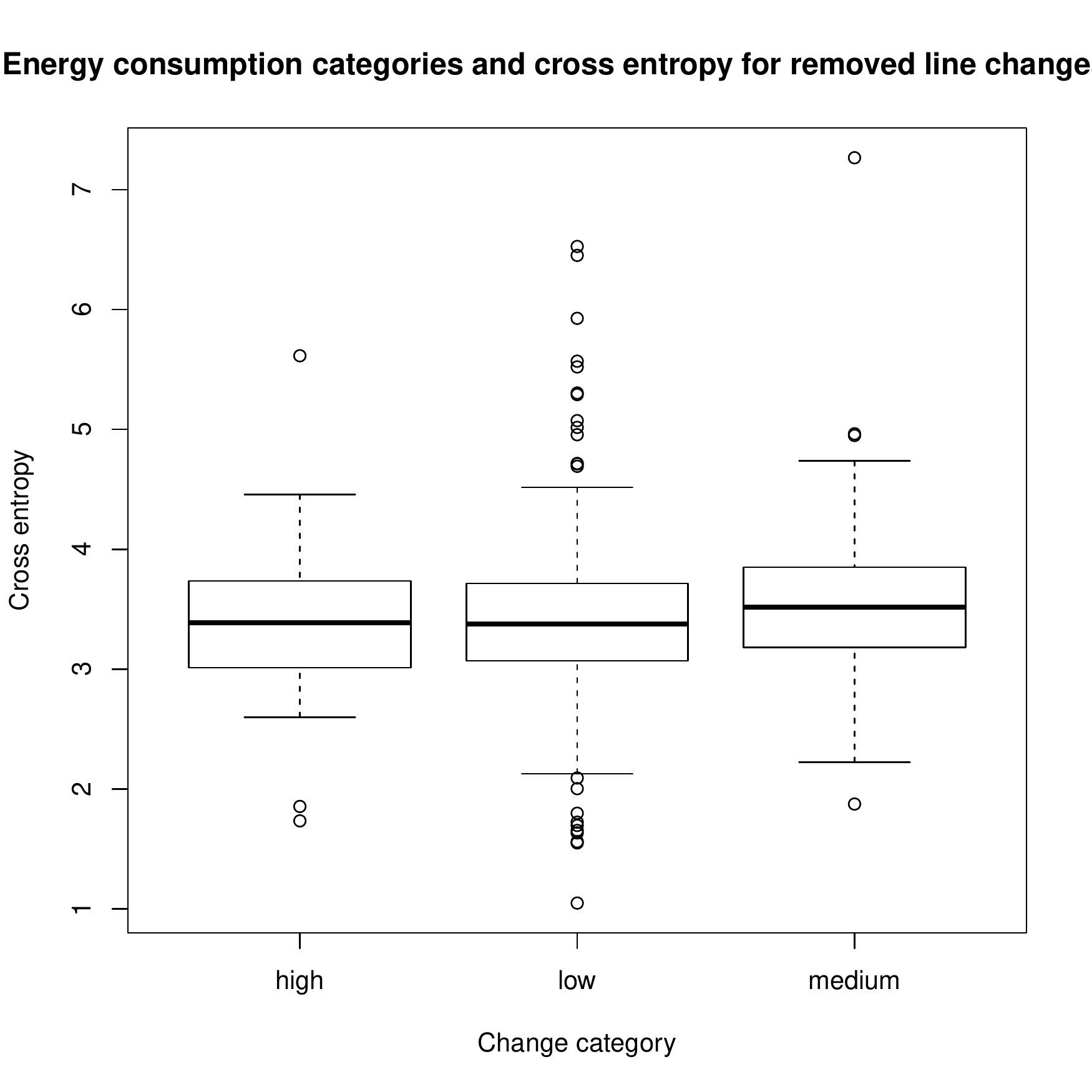}
  \caption{Changeset Cross Entropy Grouped by Change in Energy Consumption of removed line changesets}
  \label{fig:threeGroupRemoved}
\end{figure}

\paragraph{Part 3:}
Another way to check if there is a relationship between cross entropy
and energy consumption change, for a changeset, is to group the changesets
by change in energy consumption.
Figure \ref{fig:threeGroupAdded} and \ref{fig:threeGroupRemoved} show the
calculated Fennec software changeset cross entropies grouped by the changesets
change in energy consumption. A low change is considered to be -1 to 0 standard
deviations, medium is considered to be 0 to 1 standard deviations, and high
is considered to be a change greater than 1 standard deviation change in energy
consumption.

\section{Results}
Figure \ref{fig:addedLineToken} and \ref{fig:removedLineToken} show a
nonexistent or weak relationship between cross entropy and change in energy
consumption introduced by a software revision.
Figure \ref{fig:addedMovingCorp} and \ref{fig:removedMovingCorp} show a
weak relationship exists for some software corpora and their compared software
changesets; however, the majority of corpora show a weak or nonexistent
relationship between cross entropy and change in energy consumption for software
changesets with each respective corpus.
Figure \ref{fig:threeGroupAdded} and \ref{fig:threeGroupRemoved} show that the
groups by change in energy consumption all over lap. The overlap in the box
plots means that it is not easy to tell if there is a difference between
the groups. It can be noted that the median cross entropy for software changes
that introduced a high change in absolute energy consumption is lower than
the median cross entropy of software changes that introduced a low change in
absolute energy consumption.


\section{Conclusion}
Figure \ref{fig:addedLineToken} has a correlation of $0.04060102$, Figure
\ref{fig:removedLineToken} has a correlation of $-0.00239533$,
Figure \ref{fig:addedMovingCorp} has a correlation of $0.03477229$,
and Figure \ref{fig:removedMovingCorp} has a correlation of $-0.1979264$.
Each of these Pearson correlation coefficients are weak.
Based on Figure \ref{fig:addedLineToken} through \ref{fig:threeGroupRemoved}
it is unlikely that changeset cross entropy, with respect to a software's
codebase, and changeset change in energy consumption correlate.

\subsection*{Future Work}
Prior work using static analysis~\cite{muthukumarasamy2010extraction} could
be a starting point for further analysis of code changesets. The prior work
looks at the estimating energy consumption and execution time from assembly;
however, it is not simple to give developers feedback based on their code
changesets since the compiler may heavily alter the developers code.
Therefore, an approach similar to this would have to look at methods that
are applicable to uncompiled changesets. A program that performs static
analysis on changesets could be trained with existing energy profiles.

Predicting how a software change influences a software components energy
consumption and how the change to the software component affects the whole
application could be explored as well.

It may be useful to evaluate the computational difficulty of the question
``Does a given software changeset influence the average energy consumption
of a software application?'' It could also be useful to look for work with
similar problems such as detecting if a new changeset introduces a bug.

\bibliographystyle{ieeetr}
\bibliography{codeChangesetNgramsAndEnergyConsumption}
\end{document}